\documentclass[aps,preprint,nofootinbib]{revtex4}
\usepackage{graphicx, amssymb}
\begin{document}
\bibliographystyle{apsrev}

\title{Can black holes have Euclidean cores?}

\author{T.~Hirayama} 
\email{hirayama@physics.utoronto.ca}
\author{B.~Holdom} 
\email{bob.holdom@utoronto.ca}
\affiliation{
Department of Physics, 
University of Toronto, 
Toronto, Ontario M5S 1A7, Canada }


\begin{abstract}
The search for regular black hole solutions in classical gravity leads us to consider a core of Euclidean signature in the interior of a black hole. Solutions of Lorentzian and Euclidean general relativity match in such a way that energy densities and pressures of an isotropic perfect fluid form are everywhere finite and continuous. Although the weak energy condition cannot be satisfied for these solutions in general relativity, it can be when higher derivative terms are added. A numerical study shows how the transition becomes smoother in theories with more derivatives. As an alternative to the Euclidean core, we also discuss a closely related time dependent orbifold construction with a smooth space-like boundary inside the horizon.
\end{abstract}

\maketitle
\section{Introduction}
The singularity problem of general relativity (GR), as embodied by the Schwarzschild (Schd) solution, may not be a general consequence of a general classical theory of gravity. It may merely be an artifact of truncating a derivative expansion at the lowest term, the Einstein action, and restricting oneself to solutions of that truncated theory. Clearly, before the singularity at $r=0$ is reached the Einstein action can no longer be trusted. From this viewpoint it is of some interest to see whether the singularity problem that arises in classical gravity could also be resolved within classical gravity. The form of such a resolution in higher derivative theories could also have a bearing on how singularity problems are to be addressed in quantum theories.

We shall refer to some of our results as \textbf{robust} if they are likely to hold for a general classical theory of gravity with any number of derivatives, with the possible exception of isolated cases in the space of theories. These results have been found to be true for theories with various number of derivatives (including theories with up to six derivatives), and although we offer no proof, we expect such results to be true quite generally. Any particular theory involving a truncation to some number of derivatives can of course yield spurious and unphysical results (for example the ghosts and tachyons that can appear upon linearization \cite{stelle}), but the set of robust results applying to the whole set of higher derivative theories should have more significance, and should have some bearing on questions involving strong gravity.

We expect a regular (nonsingular) uncharged black hole to involve some smooth matter distribution in its interior. Interesting regular solutions must arise from physically acceptable energy-momentum tensors $T_{\mu\nu}$. In the Schd coordinate system the metric of interest is
\begin{equation}
d{s}^{2}  =  -B(r)d{t}^{2}+A(r)dr^2+r^2(d\theta^2+\sin^2\theta d\phi^2).
\end{equation}
In some other attempts to find regular black hole solutions in GR, a constraint on the metric, $A(r)B(r)=1$, is imposed and $T_{\mu\nu}$ is allowed to have whatever spatial anisotropy is needed to satisfy the equations \cite{bardeen}. But such assumptions are not correct for physical stellar solutions, and so it is questionable that they would apply to regular black holes. We are moving in the opposite direction in this work. In particular we will not constrain $A(r)$ and $B(r)$, and we will take $T_{\mu\nu}$ to have isotropic perfect fluid form. In this framework, both in GR and higher derivative theories, we will see whether various physically motivated energy conditions can be satisfied. Energy conditions play a central role in various no-go \cite{Bron} theorems in GR, but see also \cite{Barc} on the possible violation of energy conditions.

If the curvature invariants $R$, ${R}_{\mu \nu}{R}^{\mu \nu }$, ${R}_{\mu \nu \sigma \rho }{R}^{\mu \nu \sigma \rho }$... are to be regular at the origin, then this restricts the behavior of $A(r)$ and $B(r)$. For small $r$ we must have \cite{hold1}
\begin{eqnarray}\label{e3}
A( r ) & =&  1+\tilde{a}_{2}{r}^{2}+\tilde{a}_{4}{r}^{4}+...
\nonumber\\B( r ) & =& \pm 1+\tilde{b}_{2}{r}^{2}+\tilde{b}_{4}{r}^{4}+...
\end{eqnarray}
A possible multiplicative factor in $B(r)$ arising from a rescaling of $t$ is suppressed. Linear and cubic powers are not allowed. The scalar curvature, for example, at $r=0$ is $R=6(\tilde{a}_2-\tilde{b}_2)$. We find that $A(r)$ and $B(r)$ of this form provides an example of a robust solution near the origin. Note that by requiring a regular solution at $r=0$ we are assuming that the origin actually exists. Thus we do not consider the possibility that a black hole contains a wormhole, a horn or a flux tube \cite{Bron}. But in section 4 we will discuss a smooth orbifold construction that yields a black hole without a center.

In GR all the parameters in (\ref{e3}) are determined for given energy density $\rho(r)$ and pressure $p(r)$ profiles. This is not true for a general $2+4$ derivative theory, with action
\begin{equation}\label{e4}
S=\int\!{d}^{4}x\sqrt {-g}\left( R+\alpha{R}^{2}+\beta R_{{\mu\nu}}R^{\mu\nu}\right).
\end{equation}
In this case the coefficients $\tilde{a}_2$  and $\tilde{b}_2$ are free parameters. This freedom is required to match these solutions to the physical large $r$ weak gravity solutions (assuming no singularities in between). This is because the linearized $2+4$ theory has modes both increasing and decreasing exponentially with large $r$ \cite{stelle}, and the adjustment to zero of the two increasing modes requires two degrees of freedom. This ability to match the small $r$ solutions with the physical large $r$ solutions persists in even higher derivative gravity. For example in a general $2+4+6$ derivative theory (which includes terms cubic in the curvature) solutions  near $r=0$ exist for which four coefficients, $\tilde{a}_2$, $\tilde{b}_2$, $\tilde{a}_4$, $\tilde{b}_4$, are free parameters.  This freedom is required to adjust to zero the four exponentially growing modes at large $r$.

We remind the reader that the Schd solution itself is not robust since it does not exist in general gravity theories with six or more derivatives. In \cite{hold1} it was observed that for these theories, (\ref{e3}) is the only possible solution of the equations when expanded about the origin. This applies when $A(r)$ and $B(r)$ have any power law behavior near $r=0$, with smooth or vanishing matter distribution up to a possible $\delta$-function source at $r=0$.

\section{A crossing point}
The important point is that independent of what the theory is, $A(0)=1$ for a regular solution. This is a robust result.  But inside the event horizon at $r_{\rm h}$, $A(r)$ is negative, due to the change of signs of $A(r)$ and $B(r)$ at the horizon. Then to match onto a regular solution with $A(0)=1$ another sign change is required. In other words $A(r)$ must have an even number of sign changes between $r=0$ and $r=\infty$, rather than the single sign  change implied by a single horizon. Thus for a regular black hole with a center $A(r)$ must change sign at some radius $r_{*}$ which we refer to as the \textbf{crossing point}, where $r_*<r_{\rm h}$.

We shall investigate this crossing point by performing a series expansion of the equations of GR and higher derivative theories. Basic properties of this crossing point turn out to be robust. Later we shall use numerical analysis to construct solutions for all $r$ that are compatible with what we know about the origin, the crossing point, and the horizon.

As far as the matter distribution is concerned we shall consider the physically most conservative form for the energy-momentum tensor in the black hole interior: the perfect fluid form with isotropic pressure. Given that $r$ and $t$ have exchanged their space- and time-like roles for the region $r_*<r<r_{\rm h}$ we define $\rho(r)$ and $p(r)$ as follows,
\begin{equation}
T_{rr}=\rho g_{rr},\;\;\;T_{tt}=-pg_{tt},\;\;\;T_{\theta\theta}=-pg_{\theta\theta},\;\;\;T_{\phi\phi}=-pg_{\phi\phi}.
\end{equation}
This isotropic form would also arise from a general scalar field action with minimal coupling to gravity, given that the scalar field only depends on $r$. But the curvature tensor itself distinguishes between the spatial $t$ direction and the angular directions, and so the isotropic perfect fluid form could be modified through general scalar-gravity coupling terms. But this is not sufficient to justify the use of far from isotropic pressure in regions where the curvature is not large. For definiteness we discuss the case of isotropic pressure everywhere, in which case there are three equations (the $rr$, $tt$ and $\theta\theta$ equations) for the four functions $A(r)$, $B(r)$, $\rho(r)$ and $p(r)$. Our basic strategy will be to specify one of these functions and then determine the other three, but this strategy cannot guarantee that both $A(r)$ and $B(r)$ are completely regular. Nor can it guarantee that $\rho(r)$ and $p(r)$ are physically acceptable and satisfy, for example, the dominant energy condition $|p(r)|\leq\rho(r)$, or at least the weak energy condition, $\rho(r)\geq0$ and $\rho(r)+p(r)\geq0$.

We have already seen that $A(r)$ must change sign at some point, while $B(r)$ may or may not change sign at this point. Our analysis of the series expansions of the equations around the crossing point finds two types of regular solutions.
\begin{description}
\item[Type 1)] $1/A(r_{*})=0$ and $B(r_{*})= 0$
\item[Type 2)] $1/A(r_{*})=0$ and $B(r_{*})$ is finite and nonzero
\end{description}
This result is robust, and we will also find robust constraints on $T_{\mu\nu}$ for each type of solution.

The first type of solution includes the original event horizon, and so to emphasize this we replace $r_{*}$ with $r_{h}$ to discuss this case. The curvature invariants do not diverge at $r_{h}$ even in the presence of matter, but we find the following constraints, which thus apply to both the original horizon and a possible ``interior horizon''. In particular
\begin{equation}
p(r_{\rm h})=-\rho(r_{\rm h}),
\end{equation}
and the approach to the horizon from the two sides is constrained as follows,\footnote{On the side where $t$ is time-like, $T_{tt}=\rho g_{tt}$, $T_{rr}=-p_{r}g_{rr}$, $T_{\theta\theta}=-p_{\theta}g_{\theta\theta}$, $T_{\phi\phi}=-p_{\theta}g_{\phi\phi}$. The $p$ appearing in (\ref{e20}) and (\ref{e21}) is actually $p_{r}$. $p_{\theta}$ is completely smooth across the horizon.}
\begin{eqnarray}
{\rho}^{\prime}\left( r_{\rm h}\right) =-3{p}^{\prime}\left( r_{\rm h}\right) &&\mbox{\rm when $t$ is time-like,}\label{e20}\\3{\rho}^{\prime}\left( r_{\rm h}\right) =-{p}^{\prime}\left( r_{\rm h}\right) &&\mbox{\rm when $r$ is time-like.}
\end{eqnarray}
When these derivatives do not vanish, the weak and dominant energy conditions can still be satisfied in the vicinity of the horizon as long as $\rho(r_{\rm h})>0$. For the same reason $\rho(r)$ must increase as one moves away from the horizon on the time-like-$t$ side, and decrease on the time-like-$r$ side. We also note that if $p(r_{\rm h})$ and $\rho(r_{\rm h})$ and their first $n-1$ derivatives vanish then we have the following relation among their $n$th derivatives.\footnote{We don't bother to display the more general relations here.}
\begin{eqnarray}
{\rho}^{(n)}\left( r_{\rm h}\right) =-(2n+1){p}^{(n)}\left( r_{\rm h}\right) &&\mbox{\rm when $t$ is time-like}\label{e21}\\(2n+1){\rho}^{(n)}\left( r_{\rm h}\right) =-{p}^{(n)}\left( r_{\rm h}\right) &&\mbox{\rm when $r$ is time-like}
\end{eqnarray}

These robust constraints on $\rho(r)$ and $p(r)$ are acceptable for the
original horizon, where we can assume that $\rho(r)$ and $p(r)$ vanish,
but they prove to be quite restrictive for solutions in the interior of
a regular black hole where we expect a nonvanishing matter
distribution. In fact we were unable to realize a solution of this
type in our numerical analysis. Note that if anisotropic pressure is allowed then analytical solutions with an interior horizon can be found in GR \cite{bardeen}. But there are known stability problems \cite{Mye}, if not a no-go theorem \cite{Gal}, associated with this type of interior horizon.

We thus turn to the second type of regular solution where $B(r_*)\neq0$ and only one metric component is changing sign. In the situation that $A(r)$ does not change sign again before reaching $r=0$ (if it did it would have to change sign an even number of times) then $B(r)$ cannot change sign for $r<r_*$. This is because any point where $B(r)$ and not $A(r)$ changes sign has a curvature singularity. Thus for $0\leq r<r_*$, $g_{rr}$ is once again space-like and $g_{tt}$ remains space-like, and we have a metric with Euclidean signature. This will bring up some issues concerning a change in signature to which we return shortly (for other attempts to deal with this see \cite{Ellis,DeB}).

For this type of solution we do not find robust constraints on $T_{\mu\nu}$ at the crossing point. This is an indication that such solutions will be easier to realize numerically. But we do find robust constraints at the origin of the Euclidean core where
\begin{eqnarray}
p(0)&=&-\rho(0),\\
p^\prime(0)=\rho^\prime(0)&=&0,\\
{\rho}^{\prime\prime}\left( 0\right)&=&-\frac{1}{2}{p}^{\prime\prime}\left( 0\right). 
\end{eqnarray}
This shows how the geometry approaches Euclidean dS or AdS space near the origin. If $p(0)$ and $\rho(0)$ and their first $2n-2$ derivatives vanish then we have the following relation among their $(2n)$th derivatives (assuming that the higher powers in (\ref{e3}) are even).
\begin{equation}
{\rho}^{(2n)}\left( 0\right)=-\frac{1}{n+1}{p}^{(2n)}\left( 0\right) 
\end{equation}
We have seen that $p=-\rho$, or in other words $T_{\mu\nu}\propto g_{\mu\nu}$, must hold at the original horizon, and now we see that it must also hold either at the crossing if it is of type 1, or at the origin if the crossing is of type 2.

\section{Solutions close to the crossing point}
We find robust regular solutions of type 2 of the following form.
\begin{eqnarray}\label{e8}
{\frac {1}{A \left( r \right) }}&=&\sum _{i=1}^{\infty }a_{{i}}{(r-r_*)}^{i}\nonumber\\B \left( r \right)+1 &=&\sum _{i=1}^{\infty }b_{{i}}{(r-r_*)}^{i}
\end{eqnarray}
There is some number of free parameters depending on the number of derivatives in the theory.  If $p(r)$ is given then $\rho(r)$ and its derivatives at $r=r_*$ are determined.  For GR all the $a_i$ and $b_i$ coefficients are determined, for $2+4$ gravity $a_1$ and $b_1$ remain free parameters, for $2+4+6$ derivative gravity $a_1$, $b_1$, $a_2$, $b_2$ are free, etc. The number of parameters is the same as for the regular solutions at the origin in (\ref{e4}), and it is the same as the number needed to match onto the Newtonian solution at large $r$.

The existence of these robust solutions is quite remarkable since it implies that all quantities remain finite and are smoothly matched at the boundary between spaces of different signature. In particular the extrinsic curvature vanishes on both sides of the boundary. We emphasize that we are dealing with solutions of different theories, since the $\sqrt{-g}$ that appears in the action of the theory with Lorentzian signature must be replaced by $\sqrt{g}$ in the action of the theory with Euclidean signature. But after accounting for the change of sign of $A(r)$, the field equations are the same in the two theories, and this allows the solutions of the two theories to match.

One might wonder whether there is a single theory that describes both spaces including the crossing point; for example one could consider the theory with $\sqrt{-g}\rightarrow \sqrt{|g|}$. But the theory then becomes nonanalytic in the fields, and if taken seriously, would generate various $\delta$-function singularities in the field equations at the crossing point. But these singularities are not unique, since different nonanalytic extensions of the theory can give different singularities while keeping the theory on either side the same.\footnote{For example if $\epsilon$ is the sign function then $\epsilon(g)^{2}=1$ up to the existence of singular, ill-defined derivatives at $g=0$.} This is related to the controversy in the attempts to define junction conditions at a signature changing boundary (see the last reference of \cite{Ellis}). We are resigned to the fact that the theory is not well defined at the crossing point, and the field equations are ambiguous at this point. We will comment more on this at the end, but for now we will simply define $\rho(r)$ and $p(r)$ by their approach to the crossing point, using the theories on either side. It is significant then that solutions exist where all physical quantities defined in this way are finite and smooth in the vicinity of the crossing point.

For these solutions (and the more general ones below) we have the following result that is specific to GR.
\begin{eqnarray}\label{e1}
\rho \left( r_* \right) &=&{\frac {8\pi \,G}{{r_*^2}}}\nonumber\\
p \left( r_* \right) &=&{\frac {8\pi \,G \left( a_{{
1}}r_*-1 \right) }{{r_*^2}}}
\end{eqnarray}
$1/A(r)$ is changing sign such that $a_1<0$, and therefore $\rho(r_*)+p(r_*)<0$. Thus the weak (and null) energy condition cannot be satisfied in the vicinity of $r_*$ in GR. This is the result that forces us to consider higher derivative theories; higher derivatives introduce the higher order coefficients of (\ref{e8}) into the results for $\rho(r_*)$ and $p(r_*)$. It is then possible, as we explore below, that a more sensible $T_{\mu\nu}$ can exist when the crossing point is in a region of high curvature.

\section{Orbifolding instead}
\begin{figure}
\begin{center}
\includegraphics[width=10cm]{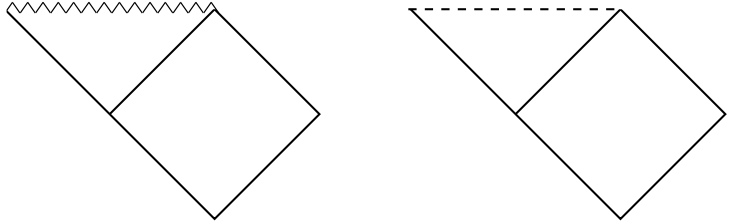}
\end{center}
\caption{We compare the Penrose diagram for a Schd black hole (left) to the orbifold construction (right). In the latter, time ends on a smooth space-like boundary rather than a singularity.}\label{f1}
\end{figure}

As an alternative to signature change, the form of the solution in (\ref{e8}) suggests a way to remove the Euclidean core region and obtain a regular black hole without a center. A smooth orbifold around $r=r_{*}$ can eliminate the core in the following way. In terms of the proper time $u$ from the crossing point, where $u^{2}=r-r_*$, the metric can be rewritten as
\begin{eqnarray}
 d{s}^{2} &=& -\widetilde{B}(u^{2})d{t}^{2}+4\widetilde{A}(u^{2})du^2
  +(u^2+r_*)^2(d\theta^2+\sin^2\theta d\phi^2)
  \nonumber\\
{\frac {1}{\widetilde{A} \left( u^{2} \right) }}&=&
  {\frac {1}{A \left( u^2+r_* \right) u^2}}=
  \sum _{i=1}^{\infty }a_{{i}}{u}^{2(i-1)}
 \nonumber \\
\widetilde{B}(u^{2})+1&=&B(u^2+r_*)+1
  =\sum _{i=1}^{\infty }b_{{i}}{u}^{2i} .
\end{eqnarray}
Recall that $a_{1}<0$ so that both $\widetilde{A}(0)$ and $\widetilde{B}(0)$ are negative. The time coordinate $u$ covers only $r-r_*\geq 0$ and the metric contains only even powers of $u$. We thus impose an orbifold condition at $u=0$ by identifying positive and negative values of $u$. This gives a smooth space-like boundary at $u=0$ where curvature invariants are clearly regular and extrinsic curvature vanishes. This would be an illustration of how a smooth time-dependent orbifold ($u$ being time-like) resolves the space-like Schd singularity, see Fig.~(\ref{f1}). Related ideas have been receiving some attention in string theory \cite{Liu}, especially in regards to the space-like cosmic singularity.

Our results concerning the weak energy condition will also apply to the orbifold construction. There is the additional question of how a gravitational collapse of a smooth matter distribution can change topology to form a black hole without a center. The orbifold possibility lies outside the main development of this paper, but we will return to it briefly in the Discussion. 

\section{More general solutions}
If the solutions in (\ref{e8}) were the most general possible for a crossing point, then we could proceed to look for completely regular solutions by numerically exploring how the regular solutions at the origin, the crossing point, and the horizon can be joined together. But we find that the solutions in (\ref{e8}) are not the most general; there are more general solutions that are not completely regular at the crossing point. Fractional powers are involved (and $\epsilon$ is the sign function).
\begin{eqnarray}\label{e7}
{\frac {1}{A\left( r\right) }}&=&\sum _{i=\{1,{3\over2},2,{5\over2}...\}}\left| r-r_*\right| ^{i}\left( a_{i}^{+}+a_{i}^{-}\epsilon(r-r_*)\right)\nonumber\\B\left( r\right)+1 &=&\sum _{i=\{{1\over2},1,{3\over2},2,{5\over2}...\}}\left| r-r_*\right| ^{i}\left( b_{i}^{+}+b_{i}^{-}\epsilon(r-r_*)\right)
\end{eqnarray}
This contains the terms in (\ref{e8}) since we can identify $a_{1}^{-}=a_{1}$, $a_{2}^{+}=a_{2}$, $a_{3}^{-}=a_{3}$, etc.\footnote{We implicitly replace $|x|\epsilon(x)$ by $x$, etc.} These solutions are robust, and it is these more general solutions that will typically emerge in the numerical analysis. We stress that despite the fractional powers, the curvature invariants (and the extrinsic curvatures) continue to remain finite as the crossing point is approached.\footnote{This is more clear if one changes the coordinate from $r$ to the proper distance from the crossing point $\propto |r-r_*|^{1/2}$.}

General values of the parameters in (\ref{e7}) can result in discontinuities in physical quantities. We will therefore study a constrained class of solutions that satisfy the following criteria.
\begin{itemize}
\item Physical quantities are continuous and are as smooth as possible at the crossing point.
\item $p(r)$ is completely regular everywhere.
\item The solutions are realizable numerically in complete black hole solutions.
\end{itemize}
The second criteria is convenient for the numerical study, since the two field equations involving $p(r)$ allow both $A(r)$ and $B(r)$ to be determined from $p(r)$ alone. The third criteria will keep us from the completely regular solutions in (\ref{e8}), at least for the time being. 

We consider GR first. Here the structure of the equations imply that a completely regular $p(r)$ implies a completely regular $1/A(r)$ of the form in (\ref{e8}). $\rho(r)$ is then at least continuous at the crossing, and to ensure that the equations are consistent with a continuous $p(r)$ requires that
\begin{equation}\label{e2}
b_{1}^{+} =-{1\over4}\left(b_{1/2}^{-2}+b_{1/2}^{+2}\right).
\end{equation}
This then implies that the curvature invariants $R$, ${R}_{\mu \nu}{R}^{\mu \nu }$ and ${R}_{\mu \nu \sigma \rho }{R}^{\mu \nu \sigma \rho }$ are also continuous. Next we must eliminate possible $| r-r_* |^{1/2}$ or $| r-r_* |^{1/2}\epsilon(r-r_*)$ terms in the equations for consistency with a regular $p(r)$. In other words $p^{\prime}(r)$ must be continuous, which requires that
\begin{equation}
b_{3/2}^{\pm}=-b_{1/2}^{\pm}{1+2b_{1}r_{*}+b_{1/2}^{+}b_{1/2}^{-}r_{*}\over3r_{*}}.
\end{equation}
Similarly demanding continuity of higher derivatives of $p(r)$ will determine $b^\pm_{{i+1/2}}$ for $i\geq 2$.

Once $A(r)$ and $B(r)$ are determined from the two $p(r)$ equations, $\rho(r)$ is determined from the third equation. The three curvature invariants are also determined. We find that these latter four quantities contain a $| r-r_* |^{1/2}\epsilon(r-r_*)$ term with coefficients $\frac{1}{2}b^-_{{1/2}}\times$ [$a_{1}/r_{*}$, $-a_{1}/r_{*}$, $-2a_{1}/r_{*}^{3}$, $a_{1}^{2}b_{1}/r_{*}$] respectively, and a $| r-r_* |^{1/2}$ term with coefficients $\frac{1}{2}b^+_{{1/2}}\times$ [$a_{1}/r_{*}$, $a_{1}/r_{*}$, $2a_{1}/r_{*}^{3}$, $-a_{1}^{2}b_{1}/r_{*}$]. Given that the $a_{i}$'s and $b_{i}$'s are determined from the two $p(r)$ equations, we conclude that there are two parameters $b^\pm_{1/2}$ remaining in the solution expanded around the crossing point. If these parameters both vanished then we would revert to the completely regular solutions in (\ref{e8}).

We now turn to the $2+4$ gravity theory. A regular $p(r)$ in this case does not imply a completely regular $1/A(r)$. To satisfy the various criteria above it turns out that we can set $b_{1/2}^{\pm}=a_{1}^{+}=b_{1}^{+}=0$. Proceeding as before we find that continuity of $p(r)$ fixes $a_{2}^{-}$ and $b_{2}^{-}$ in terms of $a^\pm_{3/2}$ and $b^\pm_{3/2}$. This then ensures that $\Box R$ is also continuous. Continuity of $p^{\prime}(r)$ then determines $a^\pm_{5/2}$ and $b^\pm_{5/2}$ in terms of $b_{3/2}^{\pm}$ and $a^\pm_{3/2}$. The coefficients of higher fractional powers are determined similarly. Thus in $2+4$ gravity, the solutions of interest are described by the six parameters $a_{1}$, $b_{1}$, $b_{3/2}^{\pm}$ and $a^\pm_{3/2}$. If the latter four vanish, then we revert to the completely regular solutions.

When we compare the solutions in GR and $2+4$ gravity, satisfying the criteria above, we see that in $2+4$ gravity the leading fractional power in the metric is $3/2$ rather than $1/2$. The same is true for $\rho(r)$, so only in $2+4$ gravity is $\rho^{\prime}(r)$ finite and continuous. Similarly $\Box R$ is not continuous in GR, but it is in $2+4$ gravity. Thus the transition to Euclidean signature has become smoother. In GR we have noticed that the relative size of $| r-r_* |^{1/2}$ and $| r-r_* |^{1/2}\epsilon(r-r_*)$ terms in $\rho(r)$ and in the curvature invariants are determined by $b^-_{1/2}/b_{1/2}^{+}$. In $2+4$ gravity each of the three curvature invariants can depend on a different combination of the same two terms, which depend in more complicated way on the four quantities $b_{3/2}^{\pm}$ and $a^\pm_{3/2}$. In $2+4+6$ derivative gravity we would expect that solutions satisfying the criteria above have $b^{\pm}_{3/2}=a^{\pm}_{3/2}=0$, and thus the curvature invariants would have a leading fractional power of $3/2$ rather than $1/2$.

Even though the curvature invariants remain finite as the crossing point is approached, the existence of fractional powers in the metric can imply $\delta$-function singularities in the curvature invariants. This can be checked independent of the field equations, which we have explained are ambiguous at $r_{*}$. The three curvature invariants have $\delta$-function singularities when $b_{1/2}^{\pm}$ are nonvanishing. These singularities are thus absent in our solution for $2+4$ gravity. But in this theory a $\delta$-function singularity would arise in $\Box R$ for example, due to the presence of $b_{3/2}^{\pm}$ and $a^\pm_{3/2}$. In $2+4+6$ gravity these would also be absent, and so on. 

Thus the basic pattern is that the leading singularities in higher derivative theories occur in higher derivative quantities, and in this way higher derivative theories become more and more smooth. This is the case as long as the higher derivative theory has a solution in which more of the coefficients in (\ref{e7}) vanish. We must now confirm that such solutions exist, by showing that the solutions we have been describing near $r_{*}$ can actually be matched onto solutions that are regular at the origin and the horizon to form a complete black hole solution (the third criteria above). For this we need a numerical treatment. We also need a numerical treatment to check that the weak energy condition can be satisfied in the $2+4$ theory. And we have already mentioned that it was the numerical study that prompted us to consider seriously the type 2 solutions, due to the difficulty of realizing type 1 solutions.

\section{Numerical study}
For the numerical study the regular solutions at the origin and at the horizon with $\rho(r_{h})=p(r_{h})=0$ provide boundary conditions. But the equations cannot be numerically integrated starting right at the horizon. One strategy is to have the interior matter distribution vanish just before the horizon is reached, so that the Schd solution can be used to supply the initial conditions at a point just inside the horizon. The matter distribution would be continuous but nonsmooth at that point. Alternatively one can use a smooth matter distribution that vanishes at the horizon and has a only a small and high order leading derivative there. Then it can be an arbitrarily good approximation to use the Schd solution to supply the initial condition at a point just inside the horizon. There is a similar numerical problem with integrating from $r=0$ in $2+4$ gravity. Here we use the series solution in (\ref{e3}) to provide initial conditions at a small but finite value of $r$. The solution then depends on the free parameters in this expansion, of which there are two in $2+4$ gravity.

For GR there are no undetermined parameters in the boundary conditions, once $r_{h}$ is given. In fact one of the equations can be solved analytically for $1/A(r)$ given $p(r)$, and the constraints on $p(r)$ can be satisfied even before $B(r)$ is found. The numerical results for $B(r)$ then confirms what we have already alluded to for GR: $b^\pm_{{1/2}}$ are in general nonzero. This eliminates the possibility of having the completely regular solutions in (\ref{e8}) in GR, at least for the form of $T_{\mu\nu}$ we are considering.

\begin{figure}
\begin{center}
\includegraphics[width=7in]{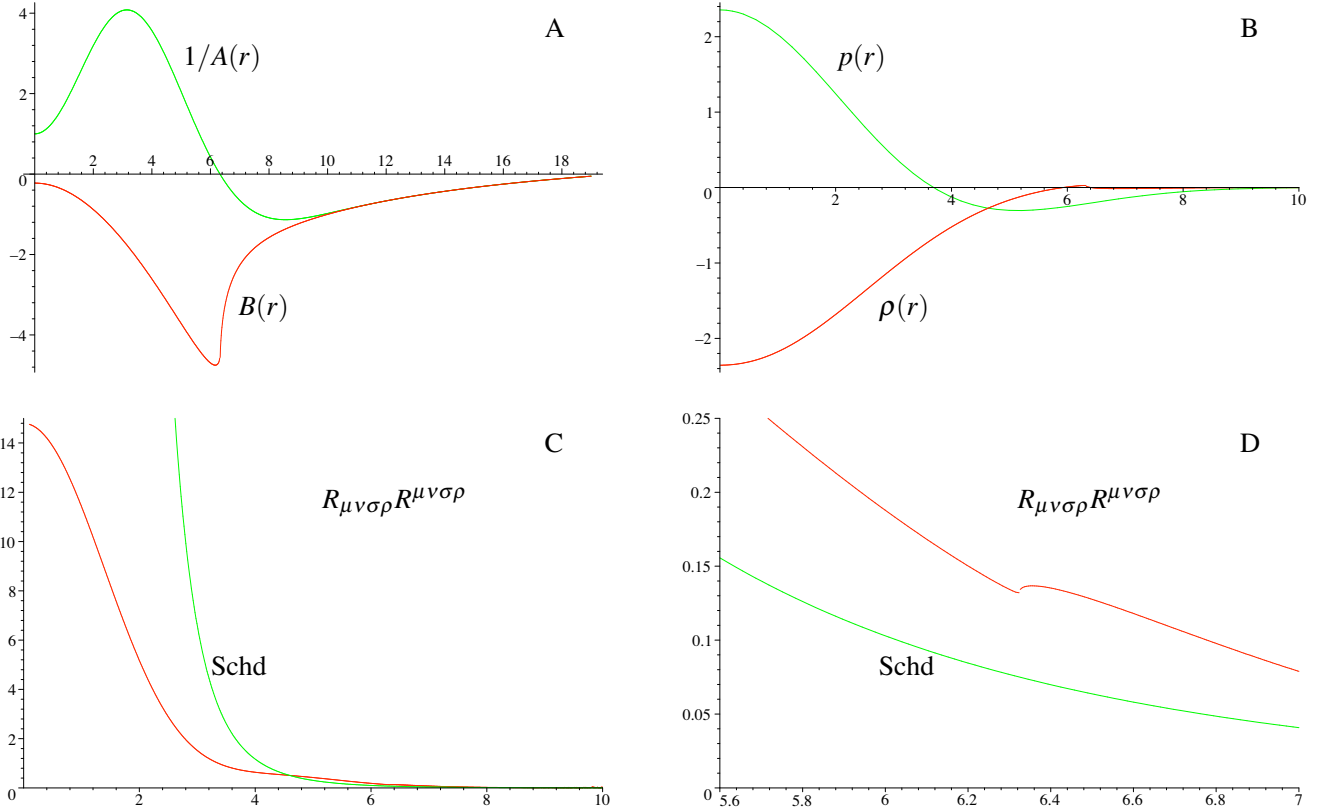}
\end{center}
\caption{A solution in GR with horizon size $r_h=20$, in Planck units. A) An infinite slope appears in $B(r)$ at the crossing $r_{*}$ where $1/A(r)$ changes sign. B) The nonsmooth behavior of $\rho(r)$ at the crossing is barely evident. C) A curvature invariant is compared to the same quantity for the Schd solution (in green). D) The region around the crossing point in (C) is enlarged.}\label{f2}
\end{figure}

\begin{figure}
\begin{center}
\includegraphics[width=7in]{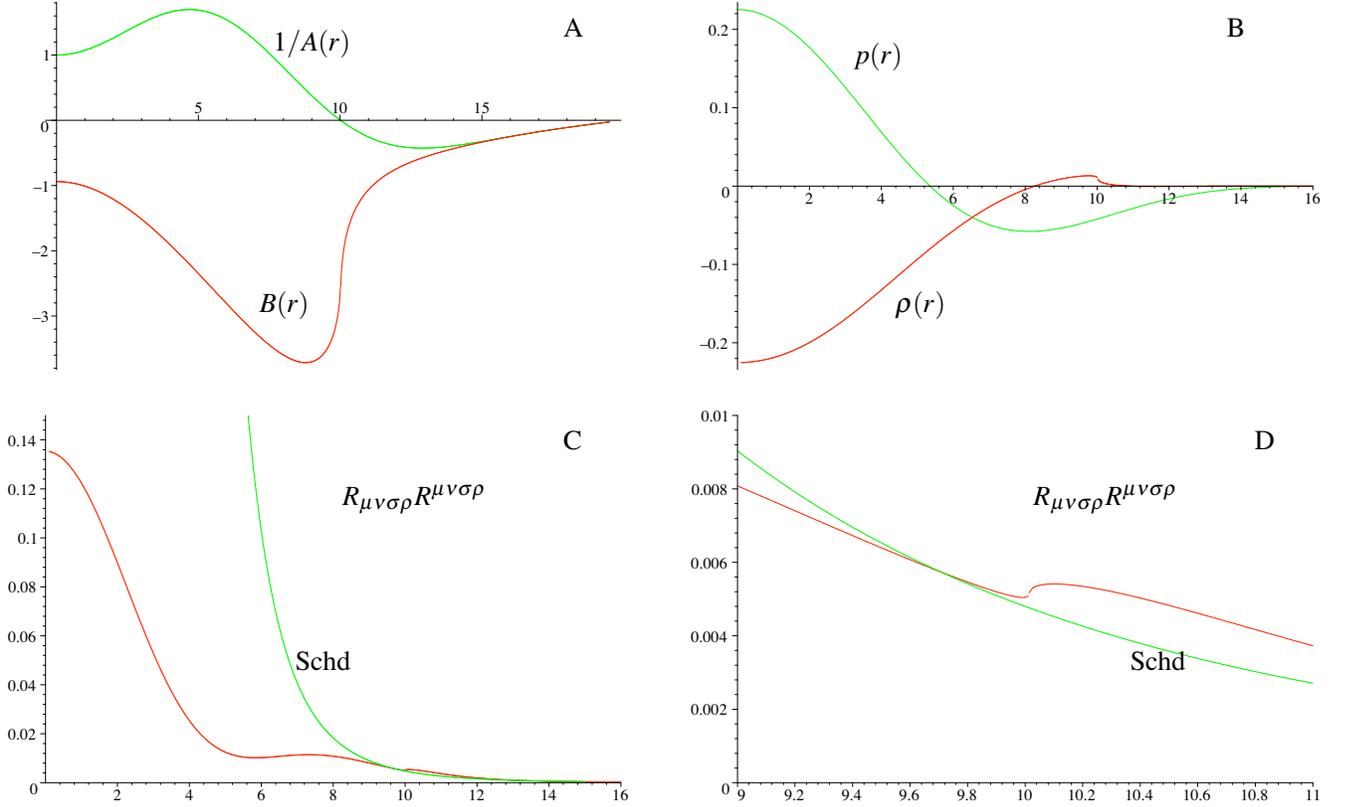}
\end{center}
\caption{Another solution in GR with horizon size $r_h=20$, differing from Fig.~(\ref{f2}) by having much smaller $\rho$ and $p$.}\label{f3}
\end{figure}

Two examples of solutions are displayed in Figures (\ref{f2}) and (\ref{f3}). It is clear that we have numerically realized the criteria of the last section, where discontinuities are absent even though $b^\pm_{1/2}\neq0$. In particular $b_{1}^{+}$ is taking on the appropriate value, although the presence of this term is not directly visible in the $B(r)$ curve. We see the presence of terms with fractional power $1/2$ in $\rho(r)$ and the curvature invariants. These terms are much more evident in $B(r)$ than in the physical quantities.  The two sets of figures also illustrate how solutions exist for a wide range in the amount of matter present, and that the matter is more concentrated in the Euclidean core when more matter is present.

\begin{figure}
\begin{center}
\includegraphics[width=7in]{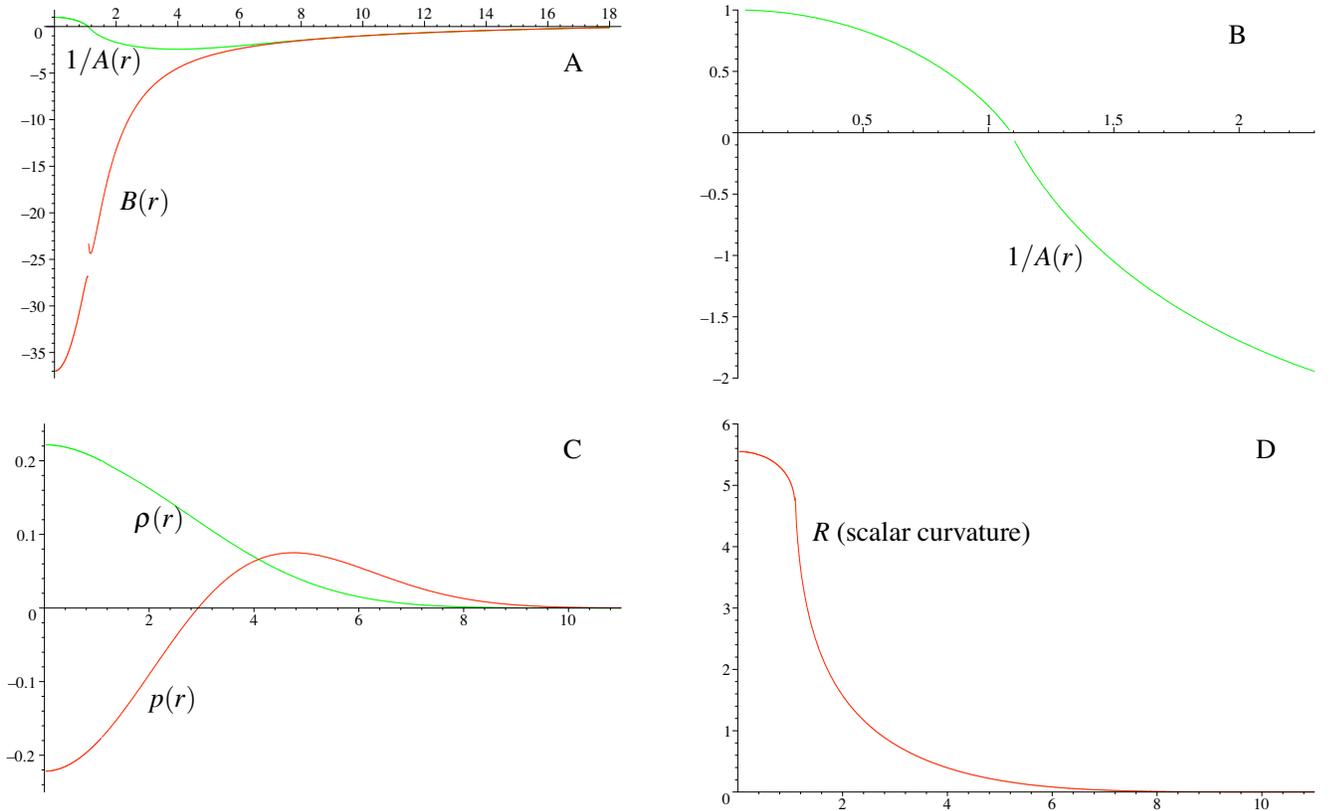}
\end{center}
\caption{A solution in the $2+4$ derivative theory of gravity with horizon size $r_h=20$, in Planck units. A) The metric functions (numerical problems are evident). B) The region around the crossing point is enlarged. C) $\rho(r)$ remains positive and it and its first derivative are continuous, unlike GR. D) The scalar curvature is continuous and finite, as in GR.}\label{f4}
\end{figure}

We turn to the $2+4$ theory. Here one must integrate the equations up from the origin and down from the horizon, and then manually adjust $p(r)$ and the free parameters $\tilde{a}_2$ and $\tilde{b}_2$ appearing in (\ref{e3}), in order obtain solutions that match in an acceptable way at the crossing point $r_*$. We verify the existence of solutions of type 2 for a range of the parameters $\alpha$ and $\beta$ in the action (\ref{e4}). An example is displayed in Fig.~(\ref{f4}) for $\alpha=-3/4$ and $\beta=1$. The results for $A(r)$ and $B(r)$ indicate the difficulties in numerically integrating the equations all the way to the crossing point, although physical quantities like $\rho(r)$ are less sensitive to the numerical problems. We find that it is possible to realize the criteria of the previous section with $b_{1/2}^{\pm}=a_{1}^{+}=b_{1}^{+}=0$. In particular we find that $\rho(r)$ and its first derivative are both finite and continuous at the crossing point, so that its leading fractional power is $3/2$ rather than $1/2$. Numerically we confirm that the curvature invariants are continuous, and in particular we find that for $R$ the $|r-r_{*}|^{1/2}\epsilon(r-r_{*})$ term dominates (see Fig.~(\ref{f4}D)) as it does for GR, while for ${R}_{\mu \nu}{R}^{\mu \nu }$ and ${R}_{\mu \nu \sigma \rho }{R}^{\mu \nu \sigma \rho }$ the $|r-r_{*}|^{1/2}$ dominates instead.

In addition to finding that the transition to a Euclidean core is smoother in $2+4$ gravity, from Fig.~(\ref{f4}) we see that the weak energy condition can be satisfied. We also see that the dominant energy condition is not satisfied in this case.  At the origin of a Euclidean core we derived earlier the general constraint $p(0)=-\rho(0)$, implying an approach to either Euclidean dS or AdS space near the origin. We find AdS for GR and dS for $2+4$ gravity. In GR we find that the value of $r_*$ is quite flexible, while in $2+4$ gravity we tend to find that $r_*$ remains small, not too much larger than the Planck length, even as $r_h$ gets large.\footnote{This assumes that $\alpha$ and $\beta$ are of order one.} In this sense the solutions of the $2+4$ theory are more restricted, although there still appears to be considerable freedom in the amount of matter in the black hole of a given radius.

We have also done some exploration in $2+4+6$ derivative theories. In these theories there are corrections to the Schd solution in the vacuum, and so the effects of matter are more difficult to disentangle. But we have confirmed the general tendency for a type 2 crossing point to form for some range of parameters in the action. There is even an indication that the crossing point can exist without the presence of matter in the region $r_{*}<r<r_{\rm h}$.

When we compare Fig.~(\ref{f4}) to Fig.~(\ref{f2}-\ref{f3}) we see that the behaviors of $\rho(r)$ and $p(r)$ are distinctly different even in regions where the curvatures are small. To understand this we consider the deviation of the metric from the Schd solution in the presence of matter, expanded about the horizon.
\begin{eqnarray}
{\frac {1}{A \left( r \right) }}&=&1-\frac{r_{\rm h}}{r}+\sum _{i=1}^{\infty }\hat{a}_{{i}}{(r-r_{\rm h})}^{i}\nonumber\\B \left( r \right) &=&1-\frac{r_{\rm h}}{r}+\sum _{i=1}^{\infty }\hat{b}_{{i}}{(r-r_{\rm h})}^{i}
\label{e5}\end{eqnarray}
For GR $\hat{b}_1$ is a free parameter, for $2+4$ gravity $\hat{b}_1$, $\hat{b}_2$, $\hat{a}_1$ are free parameters, etc. We can set these free parameters to zero to get as close as possible to the Schd solution. Then in $2+4$ gravity the deviation from the Schd solution in the presence of matter starts at higher order than it does in GR. In this case solutions to the $2+4$ gravity equations close to the horizon do not even approximately satisfy Einstein's equations, even when the horizon radius is large.\footnote{We can consider the coefficients in (\ref{e5}) assuming that the leading nonzero derivative of $p(r)$ at $r=r_h$ is $p^{(n)}(r_{\rm h})$. In GR we find that in the large $r_{\rm h}$ limit that $\hat{a}_{n+1}\propto r_{\rm h}p^{(n)}(r_{\rm h})$ and $\hat{b}_n\rightarrow0$. In $2+4$ gravity we find that $\hat{a}_{n+2}=-(2n+3)\hat{b}_{n+2}\propto r_{\rm h}^2p^{(n)}(r_{\rm h})$ for $n\geq1$. One result is that the value of Ricci scalar $R$ in $2+4$ theory is larger than that in GR when $1/r_h<|r-r_h| (\ll 1)$.} It may seem odd that higher derivative terms, whose effects should be suppressed in regions of small curvature, are able to cause a breakdown of the Einstein equations. If we let $\delta A$ and $\delta B$ denote the deviations from the Schd solution, the point is that at the horizon the derivatives of these quantities which appear in the Einstein equations are taken to vanish as initial conditions. Thus in the vicinity of the horizon the matter is influencing the higher derivatives of $\delta A$ and $\delta B$ through the 4 derivative terms. For smaller $r$ the curvature invariants gradually grow large and all terms and derivatives become of natural size.

\section{Discussion}
In conclusion we have searched for black hole solutions with a regular
center. For a regular center to exist, at least one metric component
must change sign in the interior of a black hole. Interior to this
transition, spacetime may have Lorentzian or Euclidean signature. But we
find that the Euclidean possibility is more typical and likely to
occur. The requirements on the energy momentum tensor at the crossing is
less constraining, and the numerical analysis where one integrates the
equations in from the horizon indicates that the Euclidean case is more
generic. Although the weak energy condition is not satisfied near the
crossing point in GR, it can be in higher derivative gravity.

The possibility of a regular black hole raises the issue of whether such solutions can be continuously connected to regular solutions without a black hole, where $A(r)$ and $B(r)$ are everywhere positive. At the transition between these classes of solutions the crossing point would have to merge with the horizon, so that $1/A(r)$ vanishes at $r_{h}$ but remains positive elsewhere. The horizon then becomes extremal. For the black hole with Euclidean core $B(r)$ is negative everywhere inside $r_{h}$, and thus $B(r)$ should vanish everywhere inside $r_{h}$ for the extremal solution. This allows it to continuously and smoothly connect to solutions where $B(r)$ is everywhere positive. Interestingly, just such extremal solutions have been presented for the Einstein-Yang-Mills theory in \cite{Lue}.

We have found that near the point of signature change the solution will generically
depend on fractional powers of the distance from the crossing (in the
coordinate system we have chosen). Nevertheless we have confirmed the
existence of solutions where $\rho(r)$ and $p(r)$ and various curvature
invariants are at least finite and continuous in GR. There are smoother
solutions in the $2+4$ derivative gravity theory where the leading
fractional power is higher than in GR. We expect that the trend
continues for theories with even more derivatives. Then as the number of
derivatives in the theory increases, the curvature invariants with
singular behavior will have to involve more derivatives. 

At the same time that the leading fractional powers in the metric
increase, more of the terms that are regular (the ones appearing in
(\ref{e8})) are left undetermined by the equations expanded around the
crossing point. The number of these undetermined parameters equals the
number needed to match onto the physical Newtonian solution at large
$r$. Thus we see how we are recovering the completely regular solutions
in (\ref{e8}) order by order in the expansion, as we consider theories
with more and more derivatives. The implication is that a completely
smooth transition to a Euclidean core emerges in a theory with all
derivatives, ie.~the complete derivative expansion of the true
underlying theory.

This picture is prompting us to contemplate the underlying theory. We
are also prompted by the way our description patches together Lorentzian
and Euclidean theories of classical gravity, leaving the description of
the crossing point itself ambiguous. A theory that describes a smooth
dynamical change of signature may require more degrees of freedom or
higher dimensions \cite{Mars}.  The notion that spacetime signature is
not fundamental also emerges in string theory, where it is found that a
string (M) theory on a Lorentzian torus is T-dual to a theory with a
different signature \cite{Hull}. And finally, in the spirit of quantum cosmology, the Euclidean core can perhaps be viewed as a Euclidean instanton, representing a tunnelling amplitude from the black hole interior to nothing.

In section 4 we described a smooth time dependent orbifold construction
that could realize a regular black hole without a center. Here we should
consider the more general solutions in section 5 where, for example in
GR, we have $b_{1/2}\neq 0$. In the orbifold construction this implies
that the metric depends on $|u|$, which in turn induces a
$\delta$-function in $T_{\mu\nu}$. Since there is no signature change,
this $\delta$-function now has some meaning. In other words, some kind
of brane-like source is needed at the space-like boundary. In theories
with more derivatives, where the leading fractional power is higher, it
appears that a $\delta$-function would arise in the highest derivative
terms in the field equations, again implying a $\delta$-function in
$T_{\mu\nu}$. But as before, a singularity in any given curvature
invariant would be absent in a theory with enough derivatives, meaning
that the orbifold geometry is becoming smoother. Whether a completely
smooth orbifold can be realized remains an open question.

We have labelled some of our results as robust. In particular we have
noted some properties of the crossing point, the origin of the Euclidean
core, and the event horizon which are true in GR, $2+4$ derivative
gravity, and $2+4+6$ derivative gravity. We believe that these are
examples of generic predictions of classical gravity.

\begin{acknowledgments}
This research was supported in part by the Natural Sciences and
 Engineering Research Council of Canada. T.~H.~was also supported in
 part by a JSPS Research Fellowship for Young Scientists. T.~H.~would
 like to thank K.~Hori, K.~Hosomichi and D.~Page for valuable
 conversations.
\end{acknowledgments}

\end{document}